\documentclass[9pt,twocolumn,twoside]{osajnl}
\usepackage{makecell}

\journal{ol} 

\setboolean{shortarticle}{true}

\title{Universal digital filtering for denoising volumetric retinal OCT and OCT angiography in 3D shearlet domain}

\author[1,*]{Jianlong Yang}
\author[2]{Yan Hu}
\author[1]{Liyang Fang}
\author[1]{Jun Cheng}
\author[1,2]{Jiang Liu}

\affil[1]{Cixi Institute of Biomedical Engineering, Ningbo Institute of Materials Technology and Engineering, Chinese Academy of Sciences, China}
\affil[2]{Department of Computer Science and Engineering, Southern University of Science and Technology, China}
\affil[*]{Corresponding author: yangjianlong@nimte.ac.cn}
\ociscodes{(110.4500) Optical coherence tomography; (030.4280) Noise in imaging systems; (110.4155) Multiframe image processing; (170.4470) Ophthalmology.}

\begin{abstract}
Retinal optical coherence tomography (OCT) and OCT angiography (OCTA) suffer from the degeneration of image quality due to speckle noise and bulk-motion noise, respectively. Because the cross-sectional retina has distinct features in OCT and OCTA B-scans, existing digital filters that can denoise OCT efficiently are unable to handle the bulk-motion noise in OCTA. In this Letter, we propose a universal digital filtering approach that is capable of minimizing both types of noise. Considering the retinal capillaries in OCTA are hard to differentiate in B-scans while having distinct curvilinear structures in 3D volumes, we decompose the volumetric OCT and OCTA data with 3D shearlets thus efficiently separate the retinal tissue and vessels from the noise in this transform domain. Compared with wavelets and curvelets, the shearlets provide better representation of the layer edges in OCT and the vasculature in OCTA. Qualitative and quantitative results show the proposed method outperforms the state-of-the-art OCT and OCTA denoising methods. Besides, the superiority of 3D denoising is demonstrated by comparing the 3D shearlet filtering with its 2D counterpart.    
\end{abstract}

\setboolean{displaycopyright}{true}

\begin{document}

\maketitle
Optical coherence tomography (OCT) is a non-invasive cross-sectional 3D imaging modality that has been widely used in the studies and clinical diagnosis of various diseases in ophthalmology \cite{huang1991optical}. In recent years, its functional extension, OCT angiography (OCTA) has been becoming popular because it can acquire capillary vasculature without injecting contrast agents such as fluorescent dyes \cite{gao2016optical}.\\
\indent OCT optical systems use partially coherence lasers as the light source, so the unavoidable speckle noise is a major constrain of the imaging quality \cite{schmitt1999speckle}. On the other hand, OCTA calculate the temporal decorrelation of several repeated OCT scans, which boosts the difference between vasculature and the surrounding tissue. However, both the blood flow in vessels and the involuntary eye movement of huamn lead to high decorrelation. The latter causes the bulk-motion noise in OCTA images \cite{jia2012quantitative}.
\\
\begin{figure}[b!]
\centering
\includegraphics[width=8.5cm]{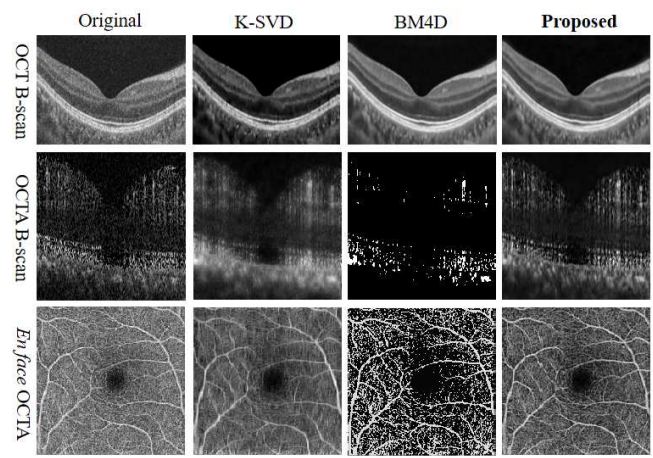}
\caption{Comparison of different methods in denoising OCT and OCTA, including the BM4D, K-SVD, and the proposed 3D shearlet filtering.}
\label{fig:fig0}
\end{figure}
\indent Minimizing the speckle noise in OCT while preserving the edge and texture information of retinal layers is very challenging. Numerous methods and algorithms have been developed for this task \cite{fang2012sparsity,cheng2016speckle,adler2004speckle,maggioni2012nonlocal,kafieh2014three}. For retinal OCTA, most of the vessels and capillaries are distributed on the plane perpendicular to the incident probe light, which makes the bulk-motion noise difficult to remove in the cross-sectional (B-scan) OCTA images. Frequency compounding has been used in denosing the bulk motion but sacrifices the axial resolution \cite{jia2012split}. Jia \textit{et al.} further used median subtraction to minimize the bulk motion noise \cite{jia2015quantitative}. 
\\
\begin{figure}[b!]
\centering
\includegraphics[width=8.5cm]{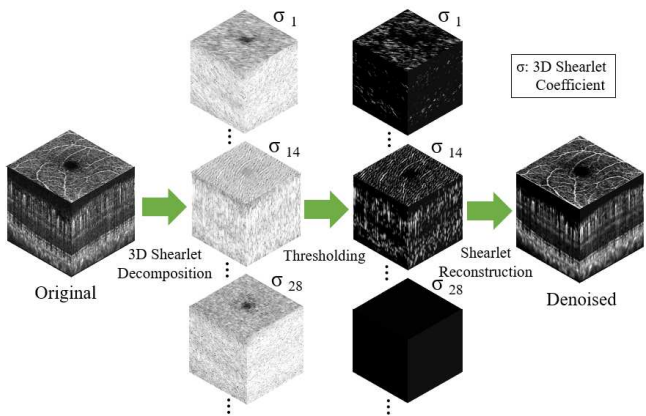}
\caption{Workflow of the proposed 3D shearlet filtering for OCT and OCTA denoising. OCTA data is employed here for demonstration. Firstly, the OCTA volume is decomposed into a set of 3D shearlet coefficients, each of them represents different features of the original data. We then apply the hard threshold to all the coefficients. Finally, all of the thresholded coefficients are used to reconstruct the denoised OCTA volume.}
\label{fig:fig1}
\end{figure}
\indent However, none of the methods above can handle both the speckle and bulk-motion noise simultaneously. As demonstrated in Fig~\ref{fig:fig0}, we employed the state-of-the-art denoising algorithm, block matching \& 4D collaborative filtering (BM4D) \cite{maggioni2012nonlocal}, and the wavelet based singular value
decomposition (K-SVD) method \cite{kafieh2014three} to denoise OCT and OCTA. The results show they could suppress the speckle noise in OCT B-scan but lead to blur and the loss of capillaries in OCTA B-scan and \textit{en face} OCTA. In this Letter, we propose to denoise both OCT and OCTA with digital filtering in shearlet domain. Shearlet is an anisotropic extension of wavelet \cite{easley2008sparse}, which has been mathematically proven to be more efficient than wavelet and curvelet in representing highly anisotropic features like edges and curvilinear structures \cite{easley2008shearlet}. Instead of using to 2D shearlets to decompose single B-scan, where the vessels and bulk-motion noise have similar point-like structures, we propose to leverage the curvilinear information of retinal vasculature in adjacent B-scans (volumetric data) by using 3D shearlet decomposition. In the 3D shearlet domain, the representations of different shearlets are capable of separating the retinal tissues and vessels with the noise efficiently. So we could minimize the noise by simply applying hard thresholding. Note that more sophisticated techniques, such as adaptive thresholding and total variation, have been widely adapted in denoising tasks \cite{adler2004speckle,easley2008shearlet}. We employ the hard thresholding by following the principle of Occam's razor, which implies using the simplest approach will justify the robustness of the proposed method to the most extent.
\\
\indent Figure~\ref{fig:fig1} is the workflow of the proposed 3D shearlet filtering for OCT and OCTA denoising. OCTA data is employed here for demonstration. The OCTA volume is decomposed into a set of 3D shearlet coefficients, each of them represents different features of the original data. For a specific B-scan which locates at the green line position, the retinal vessels and the bulk-motion noise are represented diversely by different shearlets, e.g., more noise features are emerged in $\sigma_1$ and $\sigma_{28}$ than those in $\sigma_{14}$. We then apply the hard threshold to all the coefficients. As demonstrated in Fig.~\ref{fig:fig1}, Because $\sigma_1$ and $\sigma_{28}$ has more noise features, so less information is left after the thresholding. Finally, all of the thresholded coefficients are used to reconstruct the denoised OCTA volume.  
\\
\indent Briefly, the shearlet transform is mapping a multivariate signal $f \in L^2(\mathbb{R}^2)$ to a set of coefficients $\sigma$ of a generating function $\psi$. 
\begin{equation}
f\rightarrow SH_{\psi}f = \langle f, \sigma \psi \rangle
\end{equation}
\indent The generating function $\psi$, which includes parabolic scaling variable $a>0$ for altering resolutions, shearing variable $s\in \mathbb{R} $ for altering directions, and translation variable $t\in \mathbb{R} ^2$ for altering positions, can be written as 
\begin{equation}
\psi_{a, s, t} = a^{3/4}\psi(S_s \begin{bmatrix}
a&0\\0&a^{1/2}
\end{bmatrix}(\cdot - t)) 
\end{equation}
where $S_s$ is the shearing matrix in the form of 
\begin{equation}
S_s = \begin{bmatrix}
1&s\\0&1
\end{bmatrix}.
\end{equation}
\indent We can see the shearlet transform is quite similar to the wavelet transform except it can handle anisotropic scaling and shearing. However, the direct numerical implementation of the shearlet transform is difficult. Because of the directional bias problem, the so-called cone-adapted shearlet system needs to be introduced \cite{lim2013nonseparable}. It partitions the Fourier-domain into four cones including two horizontal and two vertical high-pass region and a squared low-pass region. A scaling function $\phi$ can cover the squared region. Two new generating functions $\psi_h$ and $\psi_v$ are associated to the horizontal and vertical cones, respectively. Then the cone-adapted shearlet system $SH_{\phi,\psi_h,\psi_v}=\Phi(\phi)\cup \Psi_h(\psi_h)\cup \Psi_v(\psi_v)$ can be written as
\begin{equation}
\begin{split}
\Phi(\phi) = \{\phi_t=\phi(\cdot - t):t \in \mathbb{R}^2\}\\
\Psi_h(\psi_h) = \{\psi_h=a^{-3/4}\psi_h(A_ah^{-1} S_s^{-1}(\cdot - t))\\ :a \in (0,1], |s|\leq 1+a^{1/2}, t\in \mathbb{R}^2\}\\
\Psi_v(\psi_v) = \{\psi_v=a^{-3/4}\psi_v(A_{at}^{-1} S_s^{-T}(\cdot - t))\\ :a \in (0,1], |s|\leq 1+a^{1/2}, t\in \mathbb{R}^2\}\\
\end{split}
\end{equation}
where $A_ah = diag(a, a^{1/2})$ is the scaling matrix for the horizontal cone. $A_av = diag(a^{1/2}, a)$ is the scaling matrix for the vertical cone. The cone-adapted shearlet system can be directly digitalized by introducing a sampling factor $c = (c_1, c_2)\in \mathbb{R}_+^2$ in the translation index. 
\\
\indent The 3D digital shearlet filter is the product of two 2D digital shearlet filters in the frequency domain \cite{kutyniok2012shearlets}:
\begin{equation}
\hat{\psi}_{j,k}^{3D}(\xi)=\hat{\psi}_{j,k_1}^{2D}(\xi_1,\xi_2)\hat{\psi}_{j,k_2}^{2D}(\xi_1,\xi_3)
\end{equation} 
where $j$ and $k$ are the discrete scale and shearing parameters, respectively. $\xi_{1,2,3}$ is the 3D coordinates in Fourier domain. Thus the 3D shearlet decomposition of a signal $f\in \ell^2(\mathbb{Z}^3)$ can be defined as
\begin{equation}
DSH^{3D}_{j,k,m}(f) = (\overline{\psi_{j,k}^{3D}}\ast f)(m),
\end{equation}
where $m$ is the discrete translation parameter.
\\
\indent The proposed method was realized in MATLAB R2018a. The 3D shearlet decomposition and reconstruction of the volumetric OCT/OCTA data are implemented using the open-source ShearLab 3D code \cite{kutyniok2016shearlab}. A total of 99 shearlets are used in the representation of edges, curvilinear structures, and texture in OCT/OCTA. We employed the hard thresholding method in \cite{negi20123} as
\begin{equation}
T_{j,l} = \frac{TL\times\sigma^2}{\sigma_{j.l}},
\end{equation}
where $T_{j,l}$ and $\sigma_{j.l}$ are the threshold value and standard deviation of the shearlet coefficients in the ($j,l$)-th subband, respectively. $\sigma$ is the standard deviation of estimated noise distribution. $TL$ is the threshold level. We optimized the hard thresholding level via visual comparison of the denoised images. Throughout this Letter, We set a $\sigma$ of 30 and employed the $TL$s of 2.5 for OCT data and 1.5 for OCTA data.\\
\indent Using a personal workstation with Intel Xeon E5-2695 CPU, 128GB RAM, and Nvidia GTX 1080 Ti 12GB GPU, the processing time for a single OCT or OCTA volume is $\sim70$ s without GPU acceleration and $\sim19$ s with GPU acceleration. It also should be mentioned that the 3D shearlet filtering also has time-consumption advantage for the denoising tasks. Under the same hardware configuration, the processing time using the K-SVD and BM4D for the same OCT/OCTA volume are $\sim47$ s, and $\sim332$ s, respectively.
\\
\indent We employed TOPCON DRI OCT-1 ATLANTIS for OCT acquisition and ZEISS CIRRUS OCT with AngioPlex module for collecting OCTA data. Each of the volumetric OCT data covers $6\times6$ mm$^2$ field-of-view (FOV) corresponding to 256 B-frames. Each B-frame contains 512 A-lines and 992 pixels along the depth direction. The OCTA volume has a FOV of $3\times3$ mm$^2$. A equivalent sampling of 245 was used along the fast and slow axis directions. The OCTA data has 1024 pixels along the depth direction. The Human study protocol was approved by the Institutional Review Board of Cixi Institute of Biomedical Engineering, Chinese Academy of Sciences, and followed the tenets of the Declaration of Helsinki. Twenty healthy subjects (age $25.1\pm8.5$ years) were imaged by the machines and scan protocols mentioned above. Besides, the high-definition (HD) scan mode of the TOPCON machine, which averages 96 repeated scans at the same B-scan position, was employed as the denoised results using spatial averaging. 
\\
\indent The minimization of the OCTA bulk-motion noise using the 3D shearlet filtering is compared with two widely-used methods, median subtraction \cite{jia2015quantitative} and pixel averaging \cite{jia2012split}. The implementation of the median subtraction and pixel averaging followed the methods in \cite{jia2015quantitative} and \cite{jia2012split}, respectively. We employed a window size of 6 along the axial direction for the averaging.\\
\begin{figure}[h!]
\centering
\includegraphics[width=8.5cm]{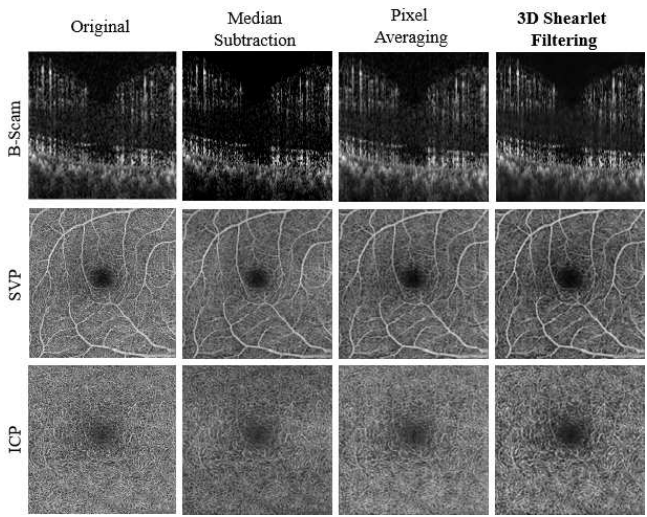}
\caption{Comparison of the denoising performance of different methods for a single OCTA volume. SVP: superficial vascular plexus. ICP: intermediate capillary plexus.}
\label{fig:fig3}
\end{figure}
\indent We evaluated the performance of denoising volumtric OCTA using the proposed method. For the clear visualization of the results in 3D, we not only demonstrated OCTA B-scans but also segmented the OCTA volume into two \textit{en face} projected slabs including superficial vascular plexus (SVP), and intermediate capillary plexus (ICP), as shown in Fig.~\ref{fig:fig3}. W used the OCT Explorer \cite{li2005optimal} for the automatic segmentation. We followed the definitions of each retinal vascular plexus in \cite{jia2015quantitative}. We can see the proposed 3D shearlet filtering provides the best vessel to noise contrast for both the B-scans and the \textit{en face} images. 
\\
\indent We further calculated the signal-to-noise ratio (SNR) of the SVP images of the twenty healthy subjects for quantitatively evaluating the performance of the OCTA denoising. We selected the SVP because it not only includes the information of both axial and transverse directions but also is widely used in the quantification of vascular biomarkers, such as foveal avascular zone (FAZ) area and vessel density \cite{gao2016optical}. We followed the method in \cite{camino2017regression} as
\begin{equation}
SNR = \frac{\overline{D}_{parafoveal}-\overline{D}_{FAZ}}{\sigma_{D_{FAZ}}},
\end{equation}
where $\overline{D}_{parafoveal}$ is the averaged flow signal in the parafovea, which is concentric with the fovea, has an outer diameter of 2.5 mm and an inner diameter of 0.6 mm. $\overline{D}_{FAZ}$ is the averaged flow signal in the FAZ and $\sigma_{D_{FAZ}}$ is its standard deviation.\\
\begin{table}[h!]
\centering
\caption{\bf Quantitative comparison of the SNR}
\begin{tabular}{p{0.4cm}cccc}
\hline
 & Original & \makecell[c]{Median \\Subtraction} & \makecell[c]{Pixel \\Averaging} & \makecell[c]{3D\\ Shearlet}\\
 \hline
SNR &  $6.67\pm1.71$ & $11.40\pm1.59$ & $12.12\pm1.56$ & $16.55\pm1.38$\\
 \hline
 \end{tabular}
\end{table}
\indent As shown Table 1, the SVP images processed by the proposed 3D shearlet filtering achieve the highest SNR in the quantitative comparison.\\
\begin{figure}[h!]
\centering
\includegraphics[width=8.5cm]{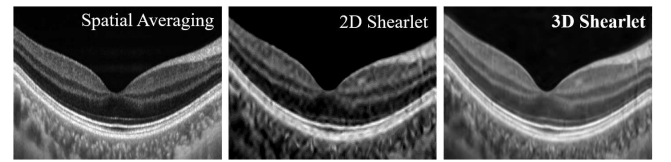}
\caption{Denoising comparison using spatial averaging, 2D shearlet filtering, and 3D shearlet filtering.}
\label{fig:fig4}
\end{figure}
\indent For evaluating the denoising performance of the proposed 3D shearlet filtering on OCT data, we also included 2D shearlet filtering in the comparison, which has been shown to be efficient in OCT denoising \cite{xu2019texture} and compressed sensing \cite{duflot2016shearlet} recently. Figure~\ref{fig:fig4} demonstrates the denoised OCT B-frame using the spatial averaging, the 2D shearlet, and the 3D shearlet proposed in this work. We can see the 3D shearlet filtering provides better SNR and resolution than its 2D counterpart, which confirms that using the additional information provided by adjacent B-frames in OCT volume could benefit the digital denoising performance \cite{cheng2016speckle}. \\
\begin{table}[h!]
\centering
\caption{\bf Quantitative comparison of the performance of the OCT denoising methods}
\begin{tabular}{cccc}
\hline
 Methods & MSE & PSNR & SSIM \\
\hline
Original &$0.291\pm0.023$ & $15.812\pm1.440$ & $0.318\pm0.055$ \\
K-SVD & $0.111\pm0.013$ & $19.931\pm0.912$ & $0.693\pm0.023$\\
BM4D & $0.075\pm0.008$ & $21.460\pm0.743$ & $0.747\pm0.015$\\
Shearlet 2D & $0.109\pm0.014$ & $20.088\pm0.959$ &  $0.703\pm0.011$\\
Shearlet 3D & $0.067\pm0.009$ & $21.813\pm0.850$ & $0.769\pm0.018$\\
\hline
\end{tabular}
  \label{tab:shape-functions}
\end{table}
\indent The quantitative evaluation of the OCT denoising performance could be conducted, under the assumption that the spatial averaging denoised image is noise-free, namely, the so-called ground truth in computer vision and image processing fields. This approach has been adopted in previous works \cite{cheng2016speckle,xu2019texture}. We employed the HD scan image in Fig.~\ref{fig:fig4} as the ground truth image $I_{gt}$. It was compared with the original or denoised images $I$. We utilized three quantitative metrics in the evaluation: mean square error (MSE), peak signal to noise ratio (PSNR), and structure similarity index (SSIM), according to their definitions in \cite{cheng2016speckle}. The results is shown in Table~\ref{tab:shape-functions}. In consistency with the qualitative comparison discussed above, the 3D shearlet filtering outperforms other OCT denoising methods on all three metrics.\\
\begin{figure}[h!]
\centering
\includegraphics[width=8.5cm]{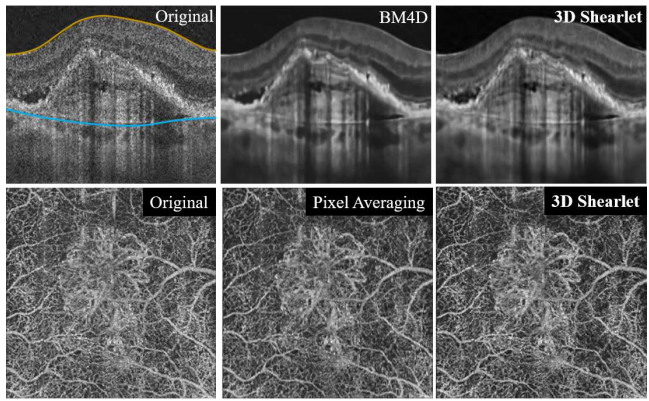}
\caption{Application of the proposed method in the OCT and OCTA denoising of a diseased case.}
\label{fig:fig5}
\end{figure}
\indent We further verified the generalization performance of the proposed method by applying it in a diseased case. It is from the pre-installed database inside the ZEISS acquisition and analysis software. The subject is 59 years old with age-related macular degeneration. The OCT and OCTA denoising results are shown in Fig.~\ref{fig:fig5} The \textit{en face} OCTA images are the axial maximum projection between the inner limiting membrane (yellow line) and the Bruch's membrane (blue line). In the OCT denoising, the 3D shearlet filtering achieves slightly better performance than the BM4D in SNR and texture preservation. It also outperform the pixel averaging in OCTA. These results are in accordance with those achieved in the healthy subjects, which may justify the effectiveness of the proposed method.\\
\indent In summary, we have demonstrated the capacity of the 3D shearlet filtering in denoising both the volumetric OCT and OCTA data. As mentioned above, shearlet is more efficient than wavelet and curvelet in representing highly anisotropic features like edges and curvilinear structures \cite{easley2008shearlet}, which explains its better performance than the wavelet-based methods, such as the K-SVD. The BM4D is not good at processing non-additive noise \cite{zhu2018temporo}, which may explain its performance degradation in handling the speckle and bulk-motion noise here. The median subtraction and pixel averaging are 1D/2D methods, which may be less efficient than 3D methods. Besides, comparing with other OCT denoising methods, the 3D shearlet is faster. Further improvement of this method could be the incorporation with the strategies that have been widely used in denoising tasks, such as local or sub-band adaptive filtering and total variation regularization. It is promising to apply the proposed method in versatile OCT/OCTA-based applications.
\vspace{2mm}
\\
\noindent\textbf{Acknowledgments.} We would like to thank the reviewers and editors for the careful reviewing and insightful comments.
\vspace{2mm}
\\
\noindent\textbf{Funding Information.} Ningbo 3315 Innovation team grant; Zhejiang Provincial Natural Science Foundation (LQ19H180001); Ningbo Public Welfare Science and Technology Project (2018C50049).
\vspace{2mm}
\\
\noindent\textbf{Disclosures.} The authors declare no conflicts of interest.


\bibliography{sample}

\bibliographyfullrefs{sample}



\end{document}